%% file: bulk.tex
\documentclass[11pt,a4paper]{article}
\usepackage{bm,jcappub,Macro,tabularx}
\usepackage{url}
\usepackage{hyperref}
\usepackage{url}
\usepackage{epstopdf}
\usepackage{enumitem}
\usepackage{amsmath}
\usepackage{colortbl}
\usepackage{color}

\input{local-macro}
\title{The Stable Clustering {\em Ansatz}, Consistency Relations and Gravity Dual of Large-Scale Structure}
\author{Dipak Munshi}
\affiliation{Astronomy Centre, School of Mathematical and Physical Sciences,\\ University of Sussex, Brighton BN1 9QH, U.K.}
\emailAdd{D.Munshi@sussex.ac.uk}
\abstract
{Gravitational clustering in the nonlinear regime remains poorly understood. Gravity dual of gravitational clustering
has recently been proposed as a means to study the nonlinear regime.
The {\em stable clustering} ansatz remains a key ingredient to our understanding of gravitational clustering in the highly nonlinear regime.
We study certain aspects of violation of the stable clustering ansatz in the gravity dual of Large Scale Structure (LSS).
We extend the recent studies of gravitational clustering using AdS gravity dual to take into 
account possible departure from the stable clustering ansatz and to arbitrary dimensions. Next, we extend the
recently introduced consistency relations to arbitrary dimensions. We use the consistency relations to test the commonly
used models of gravitational clustering including the halo models and hierarchical ans\"atze. In particular we 
establish a tower of consistency relations for the {\em hierarchical amplitudes}: $Q, R_a, R_b, S_a,S_b,S_c$ etc. as a functions of the scaled peculiar 
velocity $h$. We also study the variants of popular halo models in this context. In contrast to recent claims,
none of these models, in their simplest incarnation, seem to satisfy the consistency relations in the {\em soft} limit.}

\begin{document}
\maketitle
\flushbottom
\section{Introduction}
\label{sec:intro}
Recently completed CMB experiments, e.g. the
Planck\footnote{Planck: \href{https://www.cosmos.esa.int/web/planck}{\tt https://www.cosmos.esa.int/web/planck}} mission \citep{Planck},
have provided us with a standard model of cosmology. However, the answer to many of the outstanding questions
e.g. the nature of dark matter (DM) and dark energy (DE) as well as any possible modification
of gravity on cosmological scales remain open. Ongoing and future large scale surveys
(e.g. BOSS\footnote{Baryon Oscillator Spectroscopic Survey: 
\href{http://www.sdss3.org/surveys/boss.php}{\tt http://www.sdss3.org/surveys/boss.php}}
\citep{EW},
WiggleZ\footnote{WiggleZ Survey : \href{http://wigglez.swin.edu.au/}{\tt http://wigglez.swin.edu.au/}}
\citep{DJA},
LSST\footnote{The Large Synoptic Survey Telescope : \href{https://www.lsst.org/}{\tt http://https:/www.lsst.org/}},
DES\footnote{Dark Energy Survey: \href{http://www.darkenergysurvey.org/}{\tt http://www.darkenergysurvey.org/}}
\citep{DES},
EUCLID\footnote{EUCLID: \href{http://www.euclid-ec.org/}{\tt http://www.euclid-ec.org/}}
\citep{LAA})
will be able to answer many of these
questions. However, for the the maximum science exploitation of data from these surveys,
it will be important to understand the nature of gravitational
clustering - if possible using available analytical tools.

Many analytical techniques have been developed in recent years 
to understand nonlinear gravitational clustering, including but not limited to, 
approaches based on the renormalized perturbation theory \citep{Crocce}, 
renormalized group techniques \citep{Sabino, Floer}, Eulerian and Lagrangian 
effective field theories \citep{EFT1,EFT2,Baumann,Carraasco, DR17, LagEFT}
and more recently the time-sliced perturbation theory \citep{Blas}.
However, these approaches typically are only applicable in the
quasi-linear regime or in the intermediate regime. There are
no theoretical prescription to understand the highly nonlinear
regime of gravitational clustering. All known perturbative 
approaches and their extensions fail and numerical simulations 
are only tools to understand the highly nonlinear regime. Indeed 
few well-motivated scaling relations involving the two-point correlation functions,
and hierarchical ans\"atze, that express higher-order correlation functions in terms of two point correlation function
do exist, and provide useful clues in the highly nonlinear regime \citep{review}.

In a different context, it was recently discovered that strongly-coupled conformal field theories (CFT) can be studied 
using their weakly-coupled Anti-de Sitter (AdS) gravitational dual, also known as the AdS/CFT duality \citep{Maldacena}. The symmetries of the CFTs
manifests itself as the  symmetries of the gravitational background.
In recent years, many phenomenological applications of such duality have been suggested including the
study of quark-gluon plasma to explain the large viscosity. The idea has also
been explored in many other direction
including e.g. to understand the non-relativistic condensed matter systems \citep{Son,Bala,adams,kachru}.

The gravity dual of the large scale structure (LSS)  has recently been introduced in \citep{duality}.
In this scenario, the Universe, is pictured as a four-dimensional {\em brane} immersed in a
six-dimensional {\em bulk}. It is expected that such a formalism may allow us to understand the poorly understood highly nonlinear
gravitational clustering in terms of a weakly-coupled gravitational system in six dimensions.
It was demonstrated that the metric in six-dimension is a solution to AdS gravity coupled to a massive gauge field.
The corresponding scalar field can take the role of holographic dual of dark matter in the brane.
The scale related to the dual field is mapped into an extra radial dimension and its co-ordinate
rescaling realises the Lifshitz coefficient in the brane.

In a separate but related development, many authors in recent years have contributed to the development of {\em consistency} conditions
\citep{consistency1,Peloso,Creminelli1, Creminelli2,Valageas,const2} for gravity-induced higher-order correlation functions. 
The consistency relations are kinematic in nature. They encode correlations between large-scale linear modes and
small-scale nonlinear modes and are a direct consequence of the equivalence principle. 
They are valid despite our poor analytical understanding
of the nonlinear gravitational clustering and are unaffected by the complicated astrophysics
of star formation and supernovae feedback. This makes them particularly interesting
from an observational point of view.
These relations have recently been derived in many different context, including e.g. in redshift space \citep{const1},
as well as in the presence of primodial non-Gaussianity\citep{const4}. The density-velocity consistency relations were 
derived in \citep{const3}. The consistency relations for the CMB secondaries were investigated in \citep{Munshi1,const5}.
The consistency relations can also act as important diagnostics for detection of any departure 
from predictions of General Relativity \citep{Munshi2}. In this paper we will show how the consistency relations
can be used to constrain any analytical models of higher-order correlation hierarchy including
the halo models or the hierarchical ans\"atze.

This paper is organised as follows: In \textsection\ref{sec:Boltzmann} we outline the Lifsthiz symmetry 
of the Boltzmann equation coupled to the Poisson equation. In \textsection\ref{sec:hamilton} we review
Hamilton's scaling ansatz for gravitational clustering generalised to arbitrary dimension and without 
the assumption of stable clustering. 
The ward identities and their link to multiplet
conservation equation are discussed in \textsection\ref{sec:Ward}.
The equations for Lifshitz flow of dynamical exponents are numerically solved in \textsection\ref{sec:flow}.
The concept of consistency relation is introduced in \textsection\ref{sec:const}.
We discuss our results in \textsection\ref{sec:disc}.  
%
\section{Lifshitz symmetry and Gravity Dual of LSS}
\label{sec:Boltzmann}
%
In this section we will review the Lifshitz symmetry of the Schr\"{o}dinger space-time and their use as gravity dual of LSS.
Next, we will also discuss the  Lifshitz symmetry of the  Boltzmann-Poisson system in this context.
\subsection{Lifshitz symmetry of the Schr\"{o}dinger metric}
In the context of non-relativistic holography  Lifshitz \citep{Taylor,Nishida} and Schr\"{o}dinger \citep{Son,Bala} metrics
have recently been studied in great detail. It was shown that the 
null energy condition does not constrain the effective radius $L(r)$ (the variable $r$ denotes the radial direction) 
of the Lifshitz background. The Schr\"{o}dinger metric on the other hand 
admits monotonic behaviors of $L(r)$ due to its additional symmetry. It was shown in ref.\cite{LiuZhao1,LiuZhao2} that the null energy condition 
is sufficient to constrain the function $L(r)$ to be monotonic.
The Lifshitz and Schr\"{o}dinger metrices respectively defined in $d+2$ and $d+3$ dimensions have the following forms:
\bes
\ben
&& ds^2_{d+2}= -e^{2zr/L} dt^2 + e^{2r/L} d\bx_d^2 + dr^2 \quad  ({\rm Lifshitz}); \label{eq:Lifshitz}\\
&& ds^2_{d+3}= -e^{2zr/L} dt^2 + e^{2r/L} (2dt d\xi+d\bx_d^2) + dr^2 \quad ({\rm Schr{o}dinger}) \label{eq:sch}.
\een
\ees
The additional variable $\xi$ in the Schr\"{o}dinger metric in Eq.(\ref{eq:sch}) corresponds to the coordinate conjugate to
conserved particle number.  The variable $z$ is the dynamical exponent.
It characterizes the anisotropic temporal and spatial scaling under Lifshitz transformation: 
\ben
&& (r, \bx, t) \rightarrow (\lambda r, \lambda\bx, \lambda^z t)\quad  ({\rm Lifshitz}); \label{eq:Lifshitz1}\\
&& (r, \bx, t,\xi) \rightarrow (\lambda r, \lambda\bx, \lambda^z t, \lambda^{2-z}\xi)\quad ({\rm Schr{o}dinger}). \label{eq:sch2}
\label{eq:z}
\een
The metric given in Eq.(\ref{eq:sch}) can be generalized away from the fixed point using the following form:
\ben
ds^2_{d+3} = -e^{2A(r)}dt^2 + e^{2B(r)}(2dtd\xi +d\bx_d^2)+dr^2
\een
The Lifshitz symmetry corresponds to an isometry in the {\em bulk}. The Universe is typically associated with 
a four-dimensional {\em brane} moving in this six-dimensional bulk. The scale corresponding to the dual field theory 
gets mapped into the extra radial dimension $r$ of the six-dimensional metric. Indeed, the rescaling of the extra co-ordinate $r$
is related to the Lifshitz transformation. The above six-dimensional metric is supported by a
massive gauge field coupled to AdS gravity which is supplemented by a bulk scalar field. The bulk scalar
field plays the role of holographic dual of the dark matter in the brane.

The non-relativistic four-dimensional theories that admit anisotropic scale invariance can flow to the Lifshitz
fixed-points. The six-dimensional gravity dual theory has corresponding fixed points. 
The Lifshitz fixed-points respectively in the UV (large $r$) and IR (small $r$) reflect the
corresponding linear stages of gravitational clustering at large scale and the small scale
clustering in the highly nonlinear regime. The flow
from Lifshitz-fixed points can be studied using Renormalization Group Evolution (RGE) flows.
The flow represents the breaking of Lifshitz symmetry and correspond to the intermediate regime
of gravitational clustering. One of our aim in this paper is to establish a formal relation of the RGE description and the well known 
scaling relations of gravitational clustering.

\subsection{Lifshitz symmetry of the  Boltzmann-Poisson system}
\label{sec:Lifshitz}
\begin{figure}
\centering
\vspace{0.5cm}
{\epsfxsize=13.5 cm \epsfysize=5. cm{\epsfbox[37 380 573 591]{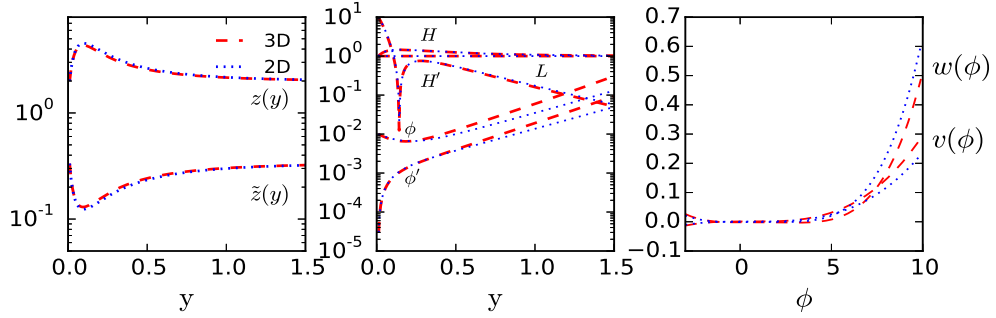}}}
\caption{The left panel shows $z$ and $\tilde{z}$ as a function of $y$. The dashed-lines correspond to 3D and 
dotted-lines to 2D. The middle panel displays $H$, $H^{\prime}$, $L$, $\phi$ and $\phi^{\prime}$ as a function
of the parameter (see Eq.(\ref{eq:eq1})-Eq.(\ref{eq:eq4}) that dictates the evolution). 
The right panel depicts $V(\phi)$ and $W(\phi)$ as a function of $\phi$.}
\label{fig:ztildez}
\end{figure}
The phase-space distribution of collisionless self-gravitating dark matter particle in an expanding background is described by
coupled Boltzmann-Poisson (or Vlasov-Poisson) equations \citep{Peebles}:
\bes
\ben
\label{eq:1}
&& {df \over d\tau} \equiv {\partial f\over \partial\tau} + {\bf u}\cdot {\bf \nabla}_{\bx} f + {d{\bf p} \over d\tau} {\bf\cdot \nabla}_{\bf p} f=0;\\
&& {d{\bf p} \over d\tau} = -am \nabla_{\bf x} \Phi(\bx,\tau);\\
\label{eq:2}
&& \triangle_{\bf x} \Phi(\bx,\tau) = 4\pi Ga^2 \bar \rho \delta(\bx,\tau); \quad \delta({\bf x}) = {\rho(\bx,\tau)-\bar\rho \over \bar\rho};
\quad \la \rho \ra = \bar\rho.
\label{eq:3}
\een
\ees
Here $f({\bf x},{\bf p},\tau)$ represents the phase space distribution, $\bx$ is the comoving coordinate, 
${\bf p}=am{\bf u}$ is the momentum, $d\bx/d\tau ={\bf u}$ is the peculiar velocity,  
$\tau$ is the conformal time, $m$ is the mass of the dark matter particles and $a$ 
corresponds to the scale factor of the Universe.
The above system admits a Lifsitz's symmetry under the following anisotropic scaling:
\ben
 \tau^{\prime} = {\lambda^{\tz}}\tau; \quad \bx^{\prime} = \lambda \bx.
\label{eq:zprime}
\een
To admit Lifsthiz symmetry [Eq.(\ref{eq:1})-Eq.(\ref{eq:3})] the following transformations need to be satisfied:
\bes
\ben
&& f^{\prime}(\bx,{\bf p}, \tau ) = f (\bx^{\prime}, {\bf p}^{\prime}, \tau^{\prime} );\\
&& \delta^{\prime}(\bx,\tau ) = \delta(\bx^{\prime} ,\tau^{\prime} );\\
&& p^{\prime}(\bx,\tau) = \lambda^{-(\tz+1)}{\bf p}(\bx^{\prime} , \tau^{\prime});\\
&& \Phi^{\prime}(\bx,\tau) = \lambda^{2(\tz-1)}\Phi(\bx^{\prime},\tau^{\prime}).
\een
\ees
As a consequence of the Lifsitz symmetry the power spectrum $P(k,\tau)$ is expressed through the following scaling function ${\cal P}$:
\bes
\ben
&& \langle \delta(\bk,\tau)\delta(\bk^{\prime},\tau)\rangle_c = \delta_{\rm 3D}(\bk+\bk^{\prime}) P(k,\tau); \\
&& P(k,\tau) = \tau^{3/\tz} {\cal P}(k/\tau^{3\tz}).
\een
\ees
Here ${\cal P}$ is an arbitrary function. The Lifsthiz exponent in the {\em brane} will be denoted by $\tilde z$
and that in the bulk will be denoted by $z$. 

A few comments about the Lifshitz symmetry are in order. Strictly speaking, Lifshitz symmetry 
is only valid in a matter dominated case. In case of two component system e.g. involving a cosmological
constant or quintessence, such a symmetry is lost. However, in the deeply non-linear regime
the dynamics is dominated by the dark matter and contributions from quintessence can be ignored,
which justifies the validity of the Lifshitz symmetry.

Indeed, it's worth pointing out that the validity of the Lifshitz symmetry does not depend on any specific value of the exponent $\tilde z$.
The exponent $z$ introduced in Eq.(\ref{eq:z}) is defined in the bulk and $\tilde z$ in Eq.(\ref{eq:zprime}) in the brane.
Typically in the perturbative regime, a fluid approximation to the Boltzmann equation is employed.
The Boltzmann-Poisson system of equation is replaced by the Continuity-Euler-Poisson system of equations.
The continuity and Euler equations are related respectively to the $0$-th and $1$-st order moments of 
the Boltzmann equation. These systems inherit the Lifshitz's symmetry from the Boltzmann equation.
However, the validity of the Boltzmann-Poisson system extends beyond the quasi-linear regime when such a {\em fluid approximation}
breaks down and shell crossing occurs. 
%
\section{Pair Conservation and Scaling Ans\"{a}tze}
\label{sec:hamilton}
%
%
We will start with the pair conservation equation and present the derivation of the two-point correlation function
in arbitrary dimension and without the usual assumption of stable cluster. These results will
be used later in the derivation of scaling exponents to generalize known results.

The BBGKY hierarchy provides an ideal set-up to study $n$-point correlation functions
of a system of self-gravitating collisionless particles in an expanding background. For the two-point correlation
function $\xi_2$ the resulting equation, related to pair conservation, has the following form in 3D \citep{Peebles}:
\ben
{\partial \bar\xi_2 \over \partial t} + {1 \over ax^2}{\partial \over \partial x}[x^2(1+\bar\xi_2)\,v]=0; \quad 
\bar\xi_2(a,x)={3 \over x^3}\int_0^x\, y^2\, dy\, {\xi}_2(a,y).
\label{eq:pair_conservation}
\een
Here, $v(a,x)$ denotes the mean relative velocity of pairs at a {\em comoving} separation $x$ and epoch $a$.
Thus, the pair conservation equation above relates the mean relative velocity of a pair of particles
and the time evolution of the correlation function in a self-gravitating system. We will introduce 
a dimensionless pair velocity $h(a,x)\equiv -v(a,x)/{\dot a} x$. Here overdot represents derivative w.r.t time. 
This equation can be simplified to the following form in an arbitrary dimension $d$:
\ben
{\partial \Xi \over \partial A } - h{\partial \Xi \over \partial X} = h\,d. 
\label{eq:pair}
\een
The following variables are used in Eq.(\ref{eq:pair}): $\Xi = \ln[1+\bar\xi_2(x,a)]; \;\; A=\ln\, a;\;\;X = \ln\, x$.
To make progress it is generally assumed $h(a,x)$ depends on $a,x$ only through $\bar\xi(a,x)$ i.e. $h(a,x)=h({\bar \xi_2}(a,x))$. 
The stable clustering ansatz corresponds to the assumption $h=1$ which amounts to assuming evolution gets frozen or {\em stabilizes} at 
smaller separation where $\bar\xi_2 \gg 1$ (equivalently $\ell \gg x$). Indeed in the limit of large separation $\bar\xi_2\ll 1$ (i.e. $\ell \approx x$), linear theory holds  
and Eq(\ref{eq:pair}) can be solved analytically. 

In general, it can be shown that a direct outcome of Eq.(\ref{eq:pair}) is that the
evolved Eulerian $\bar\xi_E(x)$ correlation function $\bar\xi_E(\ell)$ can be expressed
as a function of its linearly evolved Lagrangian counterpart $\bar\xi_L(\ell)$
but at a different length scale $\ell$ which is expressed through the
following implicit expression: 
\ben
\bar\xi_E(a,x) = u[\bar\xi_L(a,\ell)];\quad \xi_L(a,\ell)= U[\bar\xi_E(a,x)]; \quad \ell = [1+\bar\xi_E]^{1/d}x.
\een
Such an expression was first suggested in \citep{Hamilton}. Many authors have reported more accurate forms
for the fitting function $F$ (see \citep{MJW,Padmanabhan,PD96,Caldwell,PaddyOstriker,Smith03} for early results). Analytical 
progress can be made if we assume $h$ to be constant at least for a limited range of $\bar\xi_2$. In this case
a solution to Eq.(\ref{eq:pair}) takes the following form:
\ben
1+\bar\xi_2(a,x) \approx \bar\xi_2(a,x)  = a^{hd}F(a^hx)
\een
Here $F$ is an arbitrary function. The power-law index $\gamma$ can be fixed by matching 
the nonlinear $\bar\xi_2$ and the linear $\bar\xi_2 \propto a^2 x^{-(n+3)}$.
The two-point correlation function can be expressed in terms of the power-law index:
\ben
\gamma = {hd(n+d)\over h(n+d)+2}.
\label{eq:gamma_h}
\een
Now it is possible to write the two-point correlation function as:
\ben
\bar\xi_2(a,x)
 \propto a^{2hd/[2+h(n+d)]} x^{-hd(n+d)/[2+h(n+d)]}.
\label{eq:nonlinear_correlation}
\een
Numerical simulations suggests $h={2, 1}$ respectively for the {\em intermediate} and highly nonlinear regime. 
It can be shown that in general in the intermediate and highly nonlinear regime
$\bar\xi_2(a,x) \propto [\bar\xi_2(a,l)]^{hd/2}$ (see ref.\citep{MuPa} for a unified description of different regimes).
Strictly speaking, validity of such a scaling can not be established rigorously in the Fourier domain,
nevertheless, a similar scaling was found to hold in the numerical simulation \citep{Peacock}, where it takes the
following form:
\ben
&& \Delta^2(k_{NL},a) = f[\Delta^2_L(k_L,a)]; \quad  \Delta^2_{L}(k_{L},a) = F[\Delta^2_L(k_L,a)].
\een

The above expression relates the dimensionless nonlinear $\Delta^2_{NL}$ power spectrum $\Delta^2(k,\tau) = {k^3/2\pi}P(k,\tau)$ at highly nonlinear regime $k_{NL}$
as a function of linear $\Delta^2_{NL}$ power spectrum  at a linear wave-number $k_L$. The two are related through the following implicit expression:
\ben
&& k_{L} = [1 + \Delta^2_{NL}(k_{NL},a)]^{1/d}k_{NL}.
\een
In the linear regime, $\Delta^2(k_{NL},a) \le 1$, we have the following expression:
\ben
f(x)= x \,; 
\een
The intermediate and nonlinear regime is characterized by $\Delta^2(k_{NL},\tau) \ge 10$, we recover Eq.(\ref{eq:gamma_h}): 
\ben
f(x)=x^{dh/2}; 
\label{eq:scale}
\een
The inverse fitting function $F$ is uniquely defined once the function $h$ is specified \citep{NP}:
\ben
F(z) = \exp\left [ {2\over 3}\int_0^z {ds \over (1+s) h(s)}  \right ]
\een
Two possibilities are generally considered: $h=1$ and $h(n+d)={\rm const.}$

As is well known, the stable clustering approximation $h=1$ leads to an imprinting of initial conditions on the nonlinear regime 
and thus leads to an explicit break down of universality of clustering.
Indeed, the study of stable clustering and it breakdown is intimately related
to the question of the existence of {\em universal} features
in nonlinear gravitational clustering, i.e., independence of nonlinear structures of initial conditions and/or cosmological
background evolution.

Despite many years of continued investigations the validity of the stable clustering
ansatz remains disputed. Early studies in ref.\cite{PaddyOstriker} and ref.\cite{CBH} provide clear
evidence of departure from stable clustering. In ref.\citep{Jain97, Bertschinger}, on the other hand,
conclusions were drawn in favour of the stable clustering ansatz in the highly nonlinear regime.
Indeed, in more recent years, evidence was found against the stable clustering ansatz
using large simulation \citep{Smith03}. However see recent claims of validity of stable clustering \citep{Ben1,Ben2}
It is also interesting to note here that the most popular halo model based approaches
do not predict stable clustering at smaller scales \citep{halo}. A different but equivalent parametrization was introduced in  
ref.\citep{Peacock}:
\bes
\ben
&& f(x)= x^{1+\alpha}; \quad 3.5 < \alpha < 4.5; \\
&& \Delta_{NL}^2 \propto D^{(6-2\gamma)(1+\alpha)/3} k^{\gamma}; \\
&& \gamma = {3(3+n)(1+\alpha) \over 3+(3+n)(1+\alpha)}.
\een
\ees
The above expression can be recovered from Eq.(\ref{eq:scale}) by using $h/2 = (1+\alpha)/3$.
For $\alpha=1/2$ we recover the stable clustering ansatz of $h=1$ in the highly nonlinear regime.

The stable clustering ansatz amounts to assuming that once a 
collapsed object is formed, it decouples from the
cosmological expansion and stops evolving in physical co-ordinate \citep{DavisPeebles,Peebles}.
Self-similar evolution can be used to show that the two-point correlation
function takes a power-law form at small physical separation. 
When coupled to the stable clustering ansatz, it can 
be used to make exact prediction about the slope of the power law at high $k$.
Interestingly, stable clustering means the gravitational clustering at deeply
nonlinear scale remembers the initial power spectrum through its spectral index $n$.
The phenomenological approaches developed in ref.\citep{Hamilton} and ref.\citep{Peacock} are 
direct consequences of the validity of the stable clustering ansatz.

According to ref.\citep{MaFryStable} the nonlinear index $\gamma$ can be expressed 
in terms of parameters $\alpha$ and $\beta$ that define a particular halo mode e.g.  Press-Schechter (PS) the nonlinear 
power spectrum has the following form:
\ben
&&\triangle^2_{NL}(k_{NL},a) \propto k^{\gamma}; \quad \gamma = {18\beta - \alpha(n+3)\over 2(3\beta+1)}.
\een
It can be shown that assuming only the one-halo terms contribute for the power spectrum
and bispectrum, enforcing a hierarchical form for the bispectrum dictates that only possible solution  
is characterized by $\alpha =0$ and $\beta = {(3+n)/6}$. This value of $\alpha$ is not compatible
with predictions of PS mass functions that reproduces numerical results.
This result can be generalized to take into
account $h\ne 1$ in which case we get $\beta = {(3+n)h/6}$ but the conclusions remains the same.

Starting with ref.\citep{Reuamsuwan}, the generalization of stable clustering ansatz and its consequences for the
hierarchical form
of higher-order correlation function has been investigated using the BBGKY hierarchy in 
many different context \citep{YanoGouda_BBGKY}.
The connection to halo profile was investigated in \citep{YanoGouda_halo}. Many studies of stable clustering
were performed in lower dimensions 1D \citep{YanoGouda_1D}, 2D \citep{Chinag,Bagla}.
Attempts have been made in deriving stable clustering based on stability arguments \citep{stability}.
However, in numerical simulation, due to complex interplay of scale of nonlinearity, the
box size and grid size make it difficult to confirm or discard the stable clustering ansatz
with high degree of confidence. Theoretically, closure schemes of BBGKY are the only way to
study the issue but any such scheme is bound to eventually break down. 
 
\section{Ward Identities and Conservation of Multiplets}
\label{sec:Ward}
The ward identities reflect  the statistical invariance of n-point correlators $\xi_n$ 
under a infinitesimal Lifsthiz symmetry (defined as $\delta_s \tau={\tilde z}\lambda \tau $
and $\delta_s{\bf x} =\lambda{\bf x}$) transformation $\delta_s$. 
\ben
&& \xi_{1 \cdots n} \equiv \la\delta(\bx_1)\cdots \delta(\bx_n) \ra_c; \quad \delta_s \xi_{1\cdots n}=0.
\label{eq:Lif}
\een
Rotational symmetry demands that $\xi_n({\bf x}_i,\tau)\equiv \xi_n({\bf x}_{ij},\tau)$ and the Lifshitz symmetry
n-point correlators  implies \citep{consistency2} :
\bes
\ben
&& \left [\tz\tau {\partial \over \partial \tau}+ \sum_{i< j}{\bf x}_{ij}\cdot\nabla_{ij} \right ]
\xi_{1\cdots n}({\bf x}_{ij},\tau)=0.
\een
\ees
The generic solutions of Eq.(\ref{eq:Lif}) for  $\xi_n$ can be written in terms of (arbitrary) functions $F_n$ and depend on $\tau$ and $x_{ij}$ 
only through the combinations $(\tau/x^z_{ij})$ and are symmetric under the permutation $i\leftrightarrow j$. They can be written as:
\bes
\ben
&& \xi_{12}({\bf x}_1, {\bf x}_2; \tau) = F_2\left ( {\tau \over x^{\tz}_{12}}\right ); \quad
\xi_{123}({\bf x}_1, {\bf x}_2, {\bf x}_3; \tau) = F_3\left ( {\tau \over x^{\tz}_{12}},{\tau \over x^{\tz}_{13}},{\tau \over x^{\tz}_{23}}\right ).
\label{eq:tildez}
\een
\ees
By matching these results with linear predictions it is possible to fix the scaling exponent $z$.

Indeed the Ward identites and the Lifshitz symmetries are nothing but a restatement of the fact that one-point
distribution function $f(x,p,t)$ admits self-similar solution $f(x,p,t) = t^{-3-3\alpha}\hat f(x/t^{\alpha},p/t^{\beta+1/3})$
with $\beta=\alpha+1/3$ \citep{DavisPeebles}. The two-point correlation function can be expressed as a function
of arbitrary function $f$: $\xi_2 = f_2(x/t^{\alpha})$.

The Ward identites are thus a reformulation of the  
multiplet conservation equations \citep{DavisPeebles}. 
The triplet conservation equation generalizes the pair conservation equation of Eq.(\ref{eq:pair_conservation}):
\bes
\ben
&& {\partial h_{123}\over \partial \tau} + \la\nabla_{12}\cdot(h_{123}\, {\bf w}_{12,3})\ra_c+  \la\nabla_{23}\cdot(h_{123}\, {\bf w}_{23,1})\ra_c=0.
\een
\ees
We have introduced the following quantities:
\bes
\ben
&& {\bf w}_{12,3} \equiv {\la A_{123}({\bf u}_1-{\bf u}_2)\ra_c \over h_{123}}; \quad
A_{123} \equiv (1+\delta_1)(1+\delta_2)(1+\delta_3); \quad \delta_i=\delta({\bf x}_i);\\
&& h_{123} \equiv \la A_{123}\ra_c = 1+ \xi_{12}+\xi_{23}+\xi_{13}+\xi_{123}.
\een
\ees
In the highly nonlinear regime $1<<\xi_2<<\xi_3$ and if we assume ${{\bf w}_{ij,k}=-h{\cal H}{\bf x}_{ij}}$ (${\cal H} = aH$) 
that generalizes the stable clustering ansatz:
\bes
\ben
\label{eq:xi_two}
&& \xi_2({\bf x},\tau) = a^{hd}f(a^hx); \quad
\xi_3({\bf x}_1,{\bf x}_2,{\bf x}_3) = a^{2hd}f_3(a^h{\bf x}_1,a^h{\bf x}_2,a^h{\bf x}_3). 
\label{eq:xi_three}
\een
\ees
For $h=1$ we recover the well known results. By construction all hierarchical models that we will study also satisfy 
the following scaling \citep{BaSch88}:
\ben
\xi_N(\lambda{\bf x}_1, \cdots, {\lambda\bf x}_n) = \lambda^{n-1}\xi_n({\bf x}_1, \cdots, {\bf x}_n).
\een
By comparing Eq.(\ref{eq:tildez}) with expressions of $\xi_{12}$ in the linear regime $\xi_{12}=a^2x^{-(n+3)}_{12}$, and 
highly nonlinear regime [Eq.(\ref{eq:nonlinear_correlation})] we can fix the values of $\tilde z$ for the fix points:
\ben
\tilde z = {n+3 \over 4}.
\een
Thus $\tilde z$ is same both in the linear $\xi_2 \ll 1$ and highly nonlinear regime $\xi_2 \gg 1$.
In the highly nonlinear regime the $\tilde z$ do not depend on $h$. Previous results were derived
assuming stable clustering $h=1$. 
\section{Lifshitz Flow of the Dynamical Exponent}
\label{sec:flow}
%
We have identified the mapping between the Lifshitz dynamical exponent $z$ for the bulk and
that for the brane $\tz$ in quasi-linear, intermediate and highly nonlinear regime without the stable
clustering ansatz and in arbitrary dimension,  we next study the renormalization group flow 
between fixed points. The variable $r$ in the dual theory (see Eq.(\ref{eq:Lifshitz1})-Eq.(\ref{eq:sch2})) 
from an initial condition at large $r$
where the perturbations are in the linear regime to small $r$.
These correspond to UV and IR Lifshitz fixed points respectively in the dual system.

\ben
S = \int d^{d+3}{\bf x} \sqrt{-g}\left[R -2V(\phi)-{1\over 2}(\partial \phi)^2 -
{1\over 4}F_{\mu\nu}F^{\mu\nu} - {1\over 2}W(\phi)A_{\mu}A^{\mu}\right ]
\label{eq:action}
\een
The renormalized group flow is triggered and controlled by a scalar filed 
$\phi$ and potential $V(\phi)$. The coupling to the Gauge field is dictated by $W(\phi)$.
\bes
\ben
&& ds^2 = -\exp {[2A(y)]}dt^2 +\exp[2B(y)] (2dtdy + d{\bx^2}) + dy^2; \\
&&  dr^2 = \exp[-2B(y)]dy^2\\
&& \phi=\phi(y); \quad A_{\mu}= H(y)\exp{A(y)}\delta_{\mu}^0.
\een
\ees
For a $d$-dimensional space the equations for the vector and scalar fields $A_{\mu}$  and $\phi$ are \citep{LiuZhao2}:
\bes
\ben
&& \phi^{\prime\prime} +(d+2)\phi^{\prime}B^{\prime} -2 \partial_{\phi}V = 0;\\
&& A^{\prime\prime}H + A^{\prime 2}H + 2A^{\prime}H^{\prime} + dB^{\prime}(H^{\prime}+A^{\prime}H)
+ H^{\prime\prime} - WH =0.
\een
\ees
Einsten's equations can be expressed as \citep{LiuZhao2}:
\ben
&&  A^{\prime\prime}-B^{\prime\prime}+2A^{\prime 2} + (d-2)A^{\prime}B^{\prime} -dB^{\prime 2} = 0;
-{1\over 2}[(H^{\prime} + A^{\prime}H)^2 + WH^2]; \\
&& (d+1)B^{\prime\prime} + {1\over 2}\phi^{\prime 2} = 0.
\een
Using the change of variable $L(r) = {1 \over B^{\prime}(r)}$; $z(r)={A^{\prime}(r) \over B^{\prime}(r)}$ \citep{LiuZhao2}:
\bes
\ben
\label{eq:eq1}
&& ({z^{\prime}L - L^{\prime} z}){H\over L^2} +{z^2 \over L^2}H +{2z \over L}H^{\prime} + {d\over L} (H^{\prime}+ {z\over L}H)
+H^{\prime\prime}-WH=0 \, ; \\
&& {\phi^{\prime\prime}} + {d+2 \over L}\phi^{\prime} -2 \partial_{\phi} V =0 \, ;\\
&& {1 \over L^2}[z^{\prime}L + (1-z)L^{\prime} + (d+2z)(z-1)] -{1\over 2}\left[(H^{\prime} + {z\over L}H)^2 + WH^2 \right ] =0 \, ;\\
&& -{d+1 \over L^2}L^{\prime} +{1 \over 2}\phi^{\prime 2}=0.
\label{eq:eq4}
\een
\ees
The primes denote derivative w.r.t. $y$. The constraint equation takes the following form:
\ben
{(d+1)(d+2)\over 2L^2}+ V - {1\over 4}\phi^{2} =0.
\een
The fixed points correspond to the following values:
\ben
&& L(r) = L_0;\quad z(r) = z_0;\quad \phi(r) = \phi_0. \\
&& W(\phi_0) = {z_0(z_0+d)\over L_0^2}; \quad V(\phi_0) = -{(d+1)(d+2) \over 2L_0^2}; \quad \partial_{\phi}V(\phi_0)=0. 
\een
The correspondence between bulk and brane relates the two scaling exponents \citep{duality}:
\ben
{\tilde z = {1 \over 2z-1}}
\een
The bulk enhanced symmetry point can be made to corresponds to $z=2, \tilde z={1\over 3}$ with specific choice of paramters.
These fixed points are given by:
\ben
&& H=0, \quad z=-{d \over 2}; \quad H=0, \quad z=1\\
&& H^2= {1 \over L^2 W}\left(2L^2W-d \mp \sqrt{d^2+4L^2W} \right), \quad
z= -{d \over 2} \left(1\mp \sqrt{1+ {4L^2 W \over d^2}} \right).
\een
These generalizes the results derived in ref.\cite{duality} to arbitrary dimension.
We will use the following forms for the potential $V(\phi)$ and gauge coupling $W(\phi)$:
\ben
\label{eq:vphi}
&& V(\phi) = V_0 + V_1\phi + {1\over 2}V_2\phi^2 +{1\over 3!}V_3\phi^3 +{1\over 4!}V_4\phi^2(\phi-\phi_0)^2 ;  \\
&& W(\phi) = W_0 + W_1\phi + {1\over 2}W_2\phi^2 + {1\over 3!}W_3\phi^3;
\label{eq:wphi}
\een
The boundary conditions impose the following constraints on the coefficients appearing in Eq.(\ref{eq:vphi}) and Eq.(\ref{eq:wphi})
(see ref.\citep{LiuZhao2} for details):
\bes
\ben
&& V_0=-{(d+1)(d+2)\over 2L^2_{IR}}; \quad W_0 = {z_{IR}(d+z_{IR}) \over L^2_{IR}};\\
&& V_1=0; \quad  W_1=0; \\
&& V_2\phi_0^2 = -3 (d+1)(d+2)\left [ {1\over L^2_{UV}} - {1\over L^2_{IR}}\right ]; \\
&& W_2\phi_0^2 = 6\left[z_{UV}\left ({d+z_{UV}\over L^2_{UV}} \right ) -  z_{UV}\left ({d+z_{IR}\over L^2_{IR}} \right )\right];\\
&& V_3\phi_0^3 = 6(d+1)(d+2) \left [ {1\over L^2_{UV}} - {1\over L^2_{IR}}\right ];\\
&& W_3\phi_0^3 =  12\left[ z_{IR}\left ( {d+z_{IR}\over L^2_{IR}} \right ) -  z_{UV}\left ({d+z_{UV}\over L^2_{UV}} \right )\right].
\een
\ees
The values we choose for numerical studies are:  $z_{IR}=z_{UV}=2$; \;\; $L_0 = 1,\quad L_{UV} = {11L_0/10},\quad L_{IR}=L_0$ and $\phi_0=1$. 
%
\section{Consistency Relations}
\label{sec:const}
%
We will use consistency relations to test variants of halo models and various hierarchical ansatze
generally used to model higher-order correlation functions.
At the level of the bispectrum the consistency relation relates the bispectrum in the squeezed limit
with the power spectrum \citep{consistency1}:
\ben
\la\delta_{\bq}(\tau)\delta_{\bk_1}(\tau)\delta_{\bk_2}(\tau)\ra^{\prime}_{\bq\rightarrow 0} = 
P_L(q,\tau)\left [ 1-{1 \over 3}{\partial \over \partial \ln k_1} +{13 \over 21}{\partial \over \partial \ln D(a)} \right  ]P(k_1,\tau).
\label{eq:Patrick}
\een
The limit $q\rightarrow 0$ is also known as the {\em soft} limit $k_1 \approx k_2 \ll q$ and for the perturbative kernel (to be introduced 
later in \textsection\ref{subsec:halo}; see Eq.(\ref{eq:PT_BI})) both LHS and RHS
of Eq.(\ref{eq:Patrick}) reduces to $[{47/21} - {1/3}(n+3)]P_L(q)P(k_1)$ \citep{MuCo}. 

For higher-order \citep{consistency1}:
\ben
&& \la\delta_{\bq}(\tau)\delta_{\bk_1}(\tau)\cdots \delta_{\bk_n}(\tau)\ra^{\prime}_{\bq\rightarrow 0} = \nn \\
&& \quad P_L(q,\tau)\left [ 1-{1 \over 3}\sum_{i=1}^n{\partial \over \partial \ln k_i} +{13 \over 21}{\partial \over \partial \ln D_+(a)} \right]
\la\delta_{\bk_1}(\tau)\cdots \delta_{\bk_n}(\tau)\ra^{\prime}
\label{eq:Patrick}
\een
Next we will impose these constraints on models of higher-order multispectra.

Next, we will investigate the consequence of these consistency relations for lower order correlation functions
in various models of correlation hierarchy.
%
%
\subsection{Hierarchical Ansat\"{z}e}
\label{subsec:hier}
%
%
We will consider the hierarchical models prescribed by \citep{SzSz,BaSch88,cumu} as a test case
to show the predictive power of consistency relations.
We will establish the consistency relations among hierarchical amplitudes of a specific model developed in ref.\citep{BaSch88}:
\bes
\ben
\label{eq:Q3}
&& B_2(\bk_1,\bk_2,\bk_3) = Q_3(P(k_1)P(k_2)+{\rm cyc.perm.}); \\
&& B_3(\bk_1,\bk_2,\bk_3,\bk_4)= R_a(P(k_1)P(|\bk_{12}|)P(|\bk_{123}|)+ {\rm cyc.perm.})\nn \\
&& \hspace{4cm} + R_b(P(k_1)P(k_2)P(k_3) + {\rm cyc.perm.}); \\
&& B_4({\bf k_1,\cdots,k_5}) = S_a \Big [ P({k}_1)P({|\bf k}_{12}|)P({|\bf k}_{123}|)P(|{\bf k}_{1234}|)+{\rm cyc.perm.} \Big ] \nn \\
&& \hspace{3cm}+ S_b \left [ P({k}_1)P({k}_2)P(|{\bf k}_{123}|)P(|{\bf k}_{1234}|) + {\rm cyc.perm.} \right ] \nn \\
&& \hspace{3cm}+ S_c \left [ P({k}_1)P({k}_2)P({k_3})P({k_4}) + {\rm cyc.perm.} \right ].  
\een
\ees
The corresponding squeezed limits are \citep{MuCo}:
\bes
\ben
\label{eq:bi1}
&& \lim_{\bq\rightarrow 0}B_2(\bk,-\bk,\bq) = 2Q_3 P_L(q)P(k); \\
\label{eq:bi2}
&& \lim_{\bq\rightarrow 0}B_3(\bk_1,\bk_2,\bk_3,\bq) = (R_a+2R_b) P_L(q)
\left ( P(\bk_1)P(\bk_2)+{\rm cyc.perm.} \right); \nn \\
\label{eq:bi3}
&&  \lim_{\bq\rightarrow 0}B_4(\bk_1,\bk_2,\bk_3,\bq) = P_L(q)[(S_a+3S_c)[P(k_1)P(|\bk_{12}|)P(|\bk_{123}|)+ {\rm cyc.perm.})\nn \\
&& \hspace{2cm} + 2(S_b+S_c)(P(k_1)P(k_2)P(k_3) + {\rm cyc.perm.}].
\een
\ees
Notice that in the squeezed limit, multispectra of a given order, behaves as a multispectra of one order less, with its topological amplitudes
{\em renormalized} e.g. the squeezed trispectra in Eq.(\ref{eq:bi1}) is an effective bispectrum with the hierarchical amplitude $Q_3^{\prime}$
determined by $R_a$ and $R_b$ that determine the trispectra: $Q_3^{\prime}=(R_a+2R_b)P_L(q)$. Similarly, the squeezed fifth order multispectra
in Eq.(\ref{eq:bi2}) can be expressed as a trispectrum with the amplitudes redefined as $R^{\prime}_a=(S_a+3S_c)P_L(q)$ and  $R^{\prime}_b=2(S_b+S_c)P_L(q)$.

Assuming a power-law power spectrum initial power spectrum $P(k)\propto k^n$ and using the scaling relation Eq.(\ref{eq:scale})
to express $\gamma$ in terms of $n$ and finally using the consistency relation Eq.(\ref{eq:Patrick}) we arrive at:
\bes
\ben
\label{eq:three}
&& 2Q_3 = \left [ 1- {1\over 3}(\gamma-3) + {13 \over 21} {2\gamma \over (n+3)} \right ]; \\
&& 2R_a+R_b= Q_3\left [ 1- {2\over 3}(\gamma-3) + {13 \over 21} {4\gamma \over (n+3)} \right ].\\
&& S_a + 2S_b +5S_c = (4R_a + 12R_b)\left [ 1- (\gamma-3) + {13 \over 21} {6\gamma \over (n+3)} \right ].
\label{eq:fifth}
\een
\ees
Thus using Eq.(\ref{eq:Patrick}) we have established a tower of consistency relations for the correlation hierarchy.
These relations depend on the initial spectral slope of the power spectrum $n$ as well as the final
power law index of the two-point correlation function $\gamma$. Notice that even if we further 
impose the condition $h(n+3)=1$, the residual dependence on $n$ in these relations will ensure
that the memory of the initial condition is retained in the final stages of the evolution.

A simplified hierarchical model was developed in ref.\citep{SzSz} where the hierarchical amplitudes of a given
order was assumed to have identical value i.e. at the fourth order $R_a=R_b=Q_4$, and similarly at fifth order we have $S_a=S_b=S_c=Q_4$.
The corresponding consistency relations can be established using Eq.(\ref{eq:three})-Eq.(\ref{eq:fifth}) with similar 
identification. Assuming a specific for $\gamma$,
the expressions with  Eq.(\ref{eq:three})-Eq.(\ref{eq:fifth}) can be used to made quantitative predictions.

If we further assume a more specific model of hierarchical clustering where $Q_3 = \nu_2$, $R_b=\nu_2^2$, $R_a=\nu_3$,
$S_a = \nu_2^3$, $S_b = \nu_2\nu_3$ and $S_c=\nu_4$, rewriting Eq.(\ref{eq:three})-Eq.(\ref{eq:fifth}) we have: 
\bes
\ben
\label{eq:nu2}
&& \nu_2 = {1\over 2}\left [1- {1\over 3}(\gamma-3) + {13 \over 21} {2\gamma \over (n+3)} \right ];\\
&& \nu_3 = -2\nu_2^2 + \nu_2\left [ 1- {2\over 3}(\gamma-3) + {13 \over 21} {4\gamma \over (n+3)} \right ];\\
\label{eq:nu3}
&& \nu_4 = -{1\over 5}\left (\nu_2^3+2\nu_2\nu_3 \right ) +{1\over 5}(4\nu_3+12\nu_2^2)\left [ 1- (\gamma-3) + {13 \over 21} {6\gamma \over (n+3)} \right ].
\label{eq:nu4}
\een
\ees
These equations completely determine the coefficients $\nu_2$, $\nu_3$, $\nu_4$ once $h$ is specified.
Alternatively assumption of a specific form for higher-order correlation hierarchy can constrain $h$.

\begin{figure}
\centering
\vspace{0.5cm}
{\epsfxsize=14.5 cm \epsfysize=4.5 cm{\epsfbox[37 415 573 591]{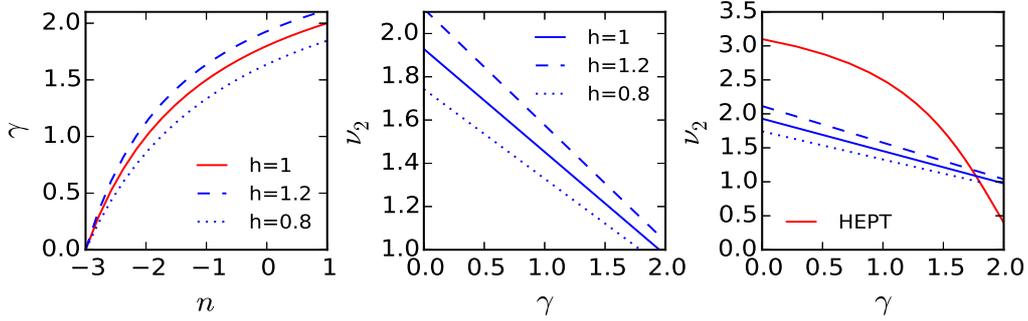}}}
\caption{The left panel shows $\gamma$ as a function of $n$.
The topological amplitude $\nu_2$ [$\nu_2 \equiv Q_3$, see Eq.(\ref{eq:nu2}) for definition, is plotted as a function of $\gamma$. 
Assuming stable clustering $h=1$ and using
consistency condition Eq.(\ref{eq:three}) gives $\nu_2=1.08$ for $\gamma=1.8$ which is remarkably close to the (phenomenological) value typically used in the
literature $\nu_2=1$.
However such agreement seems to be valid for a narrow range of (observationally interesting) spectral index $n=0$ for which $\gamma=1.8$ as can be see from
right panel where we compare numerical fits from Hyper Extened Perturbation Theory (HEPT) \citep{review} and the same predictions from the HA. The level of
agreement is rather insensitive to change in $h$.}
\label{no_before}
\end{figure}

\begin{figure}
\centering
\vspace{0.5cm}
{\epsfxsize=17.5 cm \epsfysize=4.5 cm{\epsfbox[-137 415 573 591]{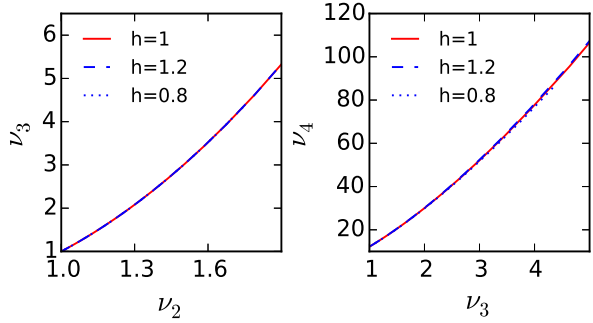}}}
\caption{The left panel shows $\nu_3$ as a function of $\nu_2$ using the consistency relation of Eq.(\ref{eq:nu3}) and the right panel
shows $\nu_4$ as a function $\nu_3$ Eq.(\ref{eq:nu4}). Different values of $h$ give nearly identical result.}
\label{fig:Q4}
\end{figure}

\begin{figure}
\centering
\vspace{0.5cm}
{\epsfxsize=13.5 cm \epsfysize=4. cm{\epsfbox[37 415 573 591]{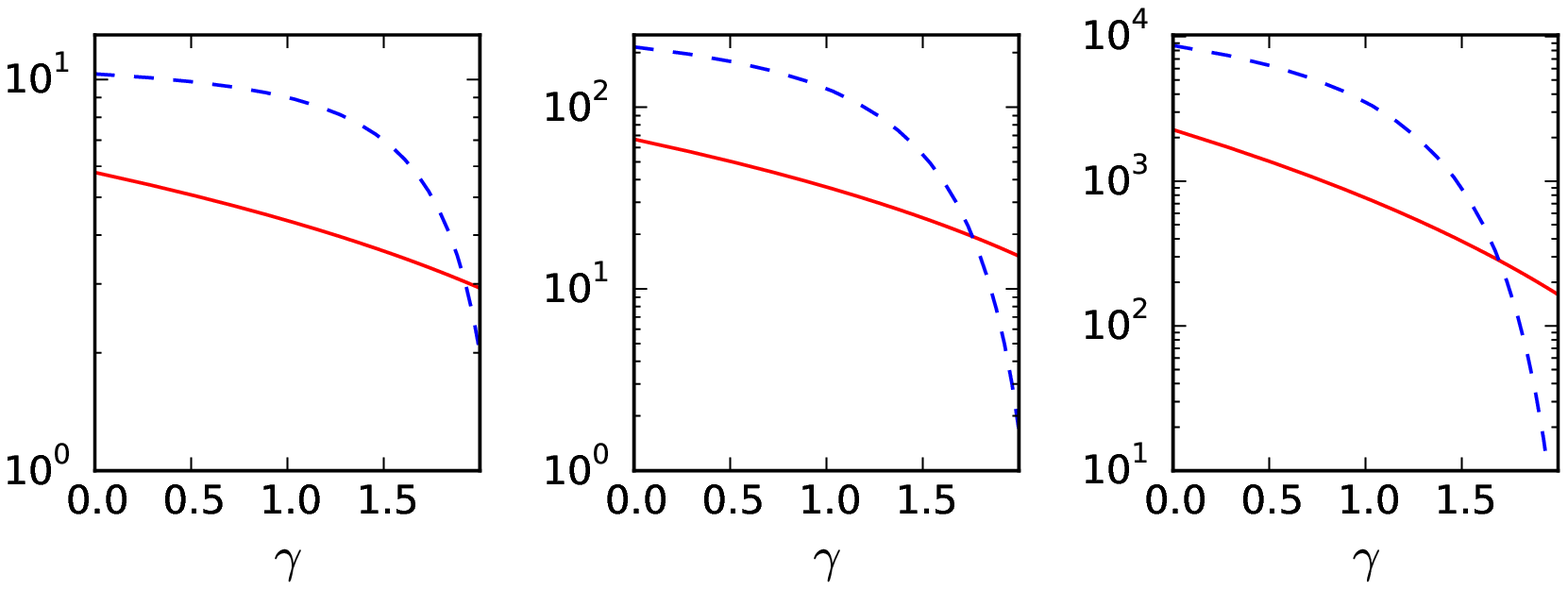}}}
\caption{We compare the predictions for lower order cumulants $S_3$ (left panel), $S_4$ (middle panel) and $S_5$ (right panel) from consistency relations and the phenomenological
fit using HEPT. Different values of $h$ give nearly identical result.}
\label{fig:comp}
\end{figure}   

\subsection{Halo models}
\label{subsec:halo}
%
In this section starting with a review of halo model we will derive the squeezed limit of halo model bispectrum.
The halo models remain the most successful in modeling gravitational clustering in the
highly nonlinear regime. Basic ingredients of the halo models include the halo profile $\rho(r)$ (or its Fourier transform $\hat\rho(k,m)$ ) \citep{halo}: 
\bes
\ben
&& \rho(r) \equiv {\rho_s \over (r/r_s)(1+r/r_s)^2 };\\
&& \hat\rho(k,m) = 4\pi \int dr r^2 \rho(r,m) {\sin(kr)\over kr}; \quad  \lim_{k\rightarrow 0}\hat \rho(m,k) \rightarrow  m;\\
&& c = {R_v \over r_s}; \quad \Delta_v=200; \quad m = {4\pi \over 3}R_v^3 \Delta_v \bar\rho.
\een
\ees
Individual halos are characterized by the mass $m$ and concentration $c(m)$. The parameters $r_s$ and $\rho_s$ can be expressed 
in terms of these parameters: 
\bes
\ben
&& r_s= \left ( {3m \over 4\pi c^3 \Delta_v \bar \rho }\right )^{1/3}; \quad \rho_s = {1\over 3}\Delta_v\,\bar\rho\,c^3 \left [\ln(1+c)-{c\over (1+c)} \right ]^{-1}.
\een
\ees
The 1-halo and 2-halo contributions  to the total power spectrum $P(k)$ from 1-halo $P_{1h}(k)$ and 2-halo terms  $P_{2h}(k)$ 
depends on the number density of halos $n(m,z)$ and $\hat\rho(k,m)$ \citep{review}:
\bes
\ben
&& P_{1h}(k) = {1\over \bar \rho^2}\, \int\, dm\, n(m,z)\, \hat\rho^2(k,m); \\
&& P_{2h}(k) = {1\over \bar \rho^2} \left [ \prod^2_{i=1} \int dm_i\, n(m_i,z)\, \hat\rho(k,m_i,z)\right ]P_{hh}(k; m_1,m_2); \\
&& P_{hh}(k ; m_1,m_2) = b_1(m_1)b_1(m_2) P_{\delta}(k);\\
&& P(k)= P_{1h}(k)+P_{2h}(k) = \epsilon_2^{[1]}(k) + [\epsilon_1^{[b_1]}(k)]^2 P_L(k).
\een
\ees
We have defined the weighted moments of the Fourier transform $\hat\rho(m,z,k)$ for an arbitrary function $\Psi(m,z)$:
\ben
\epsilon_s^{[\Psi]}(k) \equiv {1 \over \bar \rho^s} \int\, dm\, n(m)\, [\hat\rho(mk)]^s \Psi(m).
\een
The bias functions $b_i(m)$ satisfy the following relations:
\ben
&&{{1 \over \bar\rho} \int dm\, m\, n(m)=1}; \quad
{1\over \bar\rho}\int dm\, m\, n(m)\, b_i(m) = \delta_{i1}.
\een
For the halo models we will consider following parametrization \citep{review}:
\bes
\ben
&& \nu f(\nu) = A \sqrt{a\nu^2 \over 2\pi} \left [ 1 + {1 \over (a\nu^2)^p}\right ] e^{-a\nu^2 / 2}; \quad \nu = {\delta_c\over \sigma(m)}; \quad \delta_c=1.68; \\
&& b_1(\nu) = 1 + {a\nu^2 -1 \over \delta_c} + {2p \over \delta_c (1 + (a\nu^2)^p)}.
\een
\ees
For the Press-Schechter (PS) mass function 
we have $p=0$ and $q=1$. The Sheth-Tormen(ST) mass function correspond to $p=0.3,a=0.707$ and $A=0.322$.
The following biasing scheme is assumed:
\ben
\delta_h(m)\equiv \sum_s {1\over s!}b_s(m)\delta^s = b_1(m)\delta + {1\over 2} b_2(m)\delta^2 + \cdots 
\een
The second-order perturbative kernel has the following form \citep{review}:
\ben
&& B_{\rm PT}(\bk_1,\bk_2,\bk_3) = 2F_2(\bk_1,\bk_2)P_L(k_1)P_L(k_2) + {\rm cyc.perm.}; \nn \\
&& F_2(k_1,k_2) = {5 \over 7} + 
{1 \over 2} \left ({\bk_1\cdot \bk_2 \over k_1k_2} \right ) 
\left ({k_1\over k_2} +{k_2 \over k_1} \right ) + {2 \over 7} \left ({\bk_1\cdot \bk_2 \over k_1 k_2} \right )^2.
\label{eq:PT_BI}
\een
In the halo model the total bispectrum gets contributions from terms that correspond
to single, double and triple halo terms.
\ben
&& \la\delta_h(\bk_1)\delta_h(\bk_2)\delta_h(\bk_3)\ra_c \equiv B_{hhh}(\bk_1,\bk_2,\bk_3) \nn \\
&& \quad = B_{1h}(\bk_1,\bk_2,\bk_3) + B_{2h}(\bk_1,\bk_2,\bk_3) + B_{3h}(\bk_1,\bk_2,\bk_3).
\label{eq:bi_halo}
\een
The individual expressions in Eq.(\ref{eq:bi_halo}) take the following form \citep{halo}:
\bes
\ben
\label{eq:B1h}
&& B_{1h} = {1\over \bar\rho^3} \int\, dm\, n(m)\,\hat\rho(m,k_1)\hat\rho(m,k_2)\hat\rho(m,k_3).\\
\label{eq:B2h}
&& B_{2h} = {1\over \bar\rho^3}
\Big [ \int dm_1\, n(m_1\,\hat\rho(m_1,k_1)\int dm_2\; n(m_2)\,\hat\rho(m_2,k_2)\hat\rho(m_2,k_3)\Big ]\nn \\
&& \quad\quad\times P_{hh}(k_1,m_1,m_2) + \rm {cyc.perm.}\\
\label{eq:B3h}
&& B_{3h} = {1\over \bar\rho^3}\left [\prod^3_{i=1} \int dm_i\, n(m_i)\,\hat\rho(m_i,k_i) \right ]B_{hhh}(k_1,k_2,k_3; m_1,m_2,m_3).
\een
\ees
The halo bispectrum in Eq.(\ref{eq:B1h}) is related to the underlying dark matter bispectrum through the following expression:
\ben
&& B_{3h}(\bk_1,m_1; \bk_2,m_2; \bk_3,m_3) \equiv b_1(m_1)b_1(m_2)b_1(m_3)B_{\rm PT}(\bk_1,\bk_2,\bk_3) \nn \\
&& \quad + [ b_1(m_1)b_1(m_2)b_2(m_3)P_L(k_1)P_L(k_2) + {\rm cyc.perm.} ].
\een
The explicit expressions for the various terms are \citep{consistency1}:
\bes
\ben
\label{eq:1h}
&& \lim_{\bq\rightarrow 0}B_{1h}(\bk,-\bk,\bq) = {1 \over \bar \rho} \epsilon_2^{[m]}(k);\\
&& \lim_{\bq\rightarrow 0}B_{2h}(\bk,-\bk,\bq)  =  \epsilon_2^{[b_1]}(k)P_L(q);\\
&& \lim_{\bq\rightarrow 0}B_{3h}(\bk,-\bk,\bq)  =  
2\left [{13\over 14}+\left( {4\over 7}-
{1\over 2}{\partial \ln P_L \over \partial \ln k_1 } \right )\left ({\bq\cdot \bk\over q k} \right )^2 
+ {\epsilon_1^{[b_2]}(k)\over \epsilon_1^{[b_1]}(k)}
\right ]P_L(q)P_{2h}(k).\nn \label{eq:3h} \\
\een
\ees
In the limit of $k\rightarrow 0$ the 3-halo term dominates and in this limit $\epsilon_1^{[b_2]}=0$ so the
third term in Eq.(\ref{eq:3h}) do not contribute and result agrees with perturbative calculations.

Following ref.\cite{fnl_halo} we can express various contributing terms in the squeezed limit as follows:
In the limit of $k\rightarrow \infty$ the 2-halo term dominates. 
\bes
\ben
\label{eq:1hps}
&& \la\delta_{\bq}(\tau)\delta_{\bk_1}(\tau)\delta_{\bk_2}(\tau) \ra^{\prime}_{\bq\rightarrow 0}  = \la b_1\ra(k_1)P_L(q)P_{1h}(k_1); \\
&& \la b_1\ra(k) = {\int\, dm\, n(m)\,\hat\rho^2(m,k)b_1(m) \over \int\, dm\, n(m)\, \hat\rho^2(m,k)}
= {\epsilon_2^{[b_1]}(k) \over \epsilon_2^{[1]}(k)}.
\label{eq:sq1}
\een
\ees
The primes denote the fact that the vectors $\bq$, $\bk_1$ and $\bk_2$ satisfy the triangular equality.
The one-halo power spectrum  $P_{1h}$ in the nonlinear regime takes the following form \citep{MaFry}:
\bes
\ben
&&{ P_{1h}(k) \approx [D(a)]^{{6 \over n+5}-(1-2p)} k^{\gamma-3}};\label{eq:stable1}\\
&& {\gamma= {3(n+3)\over n+5} -(1-2p){3+n\over 5+n} = 2(1+p){3+n \over 5+n}}. \label{eq:stable2}
\label{eq:1halo_ps}
\een
\ees
The first term in Eq.(\ref{eq:1halo_ps}) represents the prediction of stable clustering ansatz
where as the second term corresponds to departure from it. It has the origin in the second term
of Eq.(\ref{eq:sq1}). The term doesn't vanish unless $\alpha =0$ or $n=-3$. 

Using the expressions Eq.(\ref{eq:stable1})-Eq.(\ref{eq:stable2}) in Eq.(\ref{eq:Patrick}):
\ben
&& \la\delta_{\bq}(\tau)\delta_{\bk_1}(\tau)\delta_{\bk_2}(\tau)\ra^{\prime}_{\bq\rightarrow 0} = 
\left [ 1 -{1\over 3}{(\gamma-3)} + {13 \over 21} \left ( {6 \over n+5} - (1-2p) \right )\right ]P_L(q)P_{1h}(k).\nn \\
\label{eq:sq2}
\een
Thus comparing Eq.(\ref{eq:sq1}) and Eq.(\ref{eq:sq2}) we get:
\ben
\la b_1\ra = \left [ 1 -{1\over 3}{(\gamma-3)} + {13 \over 21} \left ( {6 \over n+5} - (1-2p) \right )\right ].
\label{eq:b1}
\een
For $n=-1$ we get $\la b_1\approx 2$ which is lower than the value $\la b_1\ra \approx 3.5$ obtained by direct integration 
of the ST mass function \citep{consistency1}. A more detailed study will be presented elsewhere. 

In the presence of primordial non-Gaussianity, which we have ignored here, the $b(\nu)$ parameter in general will 
be non-local and have a $k$ dependence. 

\section{Discussions and Conclusions}
\label{sec:disc}
In this paper we have studied two recently introduced analytical methods to analyse gravitational clustering in the highly
nonlinear regime. We use the gravity dual of LSS formation to characterize the evolution of the nonlinear power spectrum.
Beyond power spectrum, we use the consistency relations in the highly nonlinear regime to test validity of
analytical models such as the halo model predictions and variants of HA. 
\begin{itemize}
\item
{\bf Evolution of Power Spectrum:} The isometry of the six-dimensional bulk manifests as Lifshitz symmetry of the Boltzmann-Poisson
equation in Eq.(\ref{eq:1})-Eq.(\ref{eq:3}) which governs the poorly understood gravitational dynamics of  LSS on the brane.
We have related the bulk Lifshitz dynamical exponent $z$ with its counterpart in the brane.
Previous results were derived using a stable clustering ansatz. We show how this assumption
can be lifted and using phenomenological scaling arguments more generalized relations can be derived in an arbitrary dimension.
In particular, we find that the exponent $z$ is independent of $h$. 
Thus the system settles in the fixed point irrespective of 
whether or not the stable clustering ansatz is violated.
We also solve Eq.(\ref{eq:eq1})-Eq.(\ref{eq:eq4}) numerically to study the RGE flow from the fixed point $z=2$ and back to $z=2$. The flow represents
the evolution of perturbations from quasi-linear regime to highly nonlinear regime through the intermediate regime.
The resulting evolution of $\tilde{z}$ (equivalently $z$) is shown in Figure -\ref{fig:ztildez}. 
We have shown that $\tilde z$ is independent of the scaled peculiar velocity $h$ and hence of the stable clustering ansatz.
In the intermediate regime $h=2$, as $\tilde z$ is independent of $z$, it implies that the self-gravitating system
is always also in a fixed point in the intermediate regime.
The evolution of $h$ - defined in Eq.(\ref{eq:pair}) - is well studied in numerical simulations
as it determines the power-law index of the nonlinear correlation function through Eq.(\ref{eq:nonlinear_correlation}).
Such a correspondence of $\tilde z$ and $h$ will be useful in building a more realistic gravity dual of LSS formation in intermediate and highly nonlinear 
regime. 

\item
{\bf Consistency Relations and Evolution of Higher-order Multispectra:}
\begin{enumerate}
\item{\bf{Hierarchal Ansatz (HA)}:}
We have used the recently derived consistency relations to put constraints on commonly used
HA, described in momentum space by Eq.(\ref{eq:bi1})-Eq.(\ref{eq:bi3}), generally used in the highly nonlinear regime to model 
higher-order correlation functions. We derive a tower of hierarchical relation
that depends on the scaled peculiar velocity $h$ through the power-law index $\gamma$
of the nonlinear two-point correlation function [see Eq.(\ref{eq:nonlinear_correlation})]. The results are derived for generic HA in Eq.(\ref{eq:three})-(\ref{eq:fifth})
and a specific model was considered in Eq.(\ref{eq:nu2})-Eq.(\ref{eq:nu4}).
In Figure -\ref{no_before} the left panel shows the $\gamma$ for different initial power law index $n$.
The middle panel shows the lowest order hierarchical amplitude $\nu_2$ as a function of $\gamma$
for various values of $h$. The comparision against HEPT is shown in the right panel.
We compare the results with HEPT predictions known to reproduce the results of numerical simulations in Figure -\ref{fig:Q4}.
In particular we show that, when the hierarchical amplitudes $\nu_n$ satisfy 
the consistency relations, they fail to reproduce numerical results. 
The HA matches with HEPT for an observationally interesting range of $\gamma\approx 1.8$.
The results are relatively insensitive to the value of $h$ chosen and do not depend on the assumption
of stable clustering. In Figure -\ref{fig:comp} we compare the predictions beyond the lowest order in non-Gaussianity. 
Indeed, while variants of HA provide useful toy-models for gravitational clustering, it is important
to realize that in the squeezed limit they do not reproduce the perturbative results.
\item{\bf{Halo Models}:}
We have also used the more realistic halo model in \textsection\ref{subsec:halo}. 
We have derived the bispectrum of the popular halo model. The bispectrum gets contribution
from single, double and triple-halo terms. Previous studies have pointed out
using theoretical arguments as well as using the numerical results that
the two-halo term dominates in the squeezed limit.
By using the same arguments we re-derive the squeezed limit bispectrum in the halo model.
However, our results do not match with that presented in ref.(\citep{consistency1}). 
In contrast to results presented in ref.(\citep{consistency1}) our results 
in the limit of $p=0$, do recover the correct the stable-clustering results. 
However, we find that with this correction, the halo model (PS) no longer satisfies the lowest order consistency relation.
More detailed analysis both analytical as well as numerical is required
to investigate this intriguing result. In particular, the result we present here only corresponds to $z=0$
and the spectral index at which the pre-factor $\la b_1\ra$  appearing in squeezed limit 
bispectrum is evaluated for $n=-1$. Indeed, more detailed study is required
to check the sensitivity of the
lower limits of the mass of halos that are included in the computation of the integrals in Eq.(\ref{eq:b1}).
Other variants of mass functions or different formulations may provide a better agreement with
the consistency relations. 
\end{enumerate}
\end{itemize}
\section*{Acknowledgements}
\label{acknow}
The author acknowledges support from the Science and Technology
Facilities Council (grant numbers ST/L000652/1).
It's a great pleasure for the author to acknowledges useful discussions with members of the
University of Sussex cosmology group. The author would like to thank Jonathan Loveday for 
for his help and suggestions to improve the draft. The author also acknowledges useful discussion
with Antonio Riotto.
\bibliography{bulk.bbl}
\end{document}

%% file: local-macro.tex
\def\la{\langle}
\def\ra{\rangle}

\def\be{\begin{equation}}
\def\ee{\end{equation}}
\def\ben{\begin{eqnarray}}
\def\een{\end{eqnarray}}
\def\nn{\nonumber}

\def\bk{{\bf k}}

\def\bk{{\bf k}}

\def\bx{{\bf x}}
\def\2p{{(2\pi)^2}}

\def\be{\begin{equation}}
\def\ee{\end{equation}}
\def\beq{\begin{equation}}
\def\eeq{\end{equation}}
\def\ben{\begin{eqnarray}}
\def\een{\end{eqnarray}}
\def\bes{\begin{subequations}}
\def\ees{\end{subequations}}

\def\nn{{\nonumber}}

\newcommand{\beqa}{\begin{eqnarray}}
\newcommand{\eeqa}{\end{eqnarray}}

\def\ikap0{{\cal J}_{\theta_0}(r)}

\def\one1{\langle \kappa_{(i)}\kappa_{(j)} \rangle}
\def\one{{[\bar \xi^{(ij)}]}}

\def\ba{\begin{eqnarray}}
\def\ea{\end{eqnarray}}

\def\bk{{\bf k}}
\def\bq{{\bf q}}

\def\tz{\tilde z}